\documentclass[twocolumn,showpacs,superscriptaddress,aps,prb]{revtex4}

\usepackage{graphicx}

\begin{document}

\title{Effect of a magnetic field on long-range magnetic order in
       stage-4 and stage-6 superconducting La$_2$CuO$_{4+y}$}
\author{B. Khaykovich}
\affiliation{Department of Physics and Center for Material
     Science and Engineering, Massachusetts Institute of Technology,
     Cambridge, MA 02139}
\author{R. J. Birgeneau}
\affiliation{Department of Physics, University of Toronto,
     Toronto, Ontario M5S 1A7, Canada}
\author{F. C. Chou}
\affiliation{Department of Physics and Center for Material
     Science and Engineering, Massachusetts Institute of Technology,
     Cambridge, MA 02139}
\author{R. W. Erwin}
\affiliation{Center for Neutron Research, NIST, Gaithersburg,
     MD 20899-8562}
\author{M. A. Kastner}
\affiliation{Department of Physics and Center for Material
     Science and Engineering, Massachusetts Institute of Technology,
     Cambridge, MA 02139}
\author{S.-H. Lee}
\affiliation{Center for Neutron Research, NIST, Gaithersburg,
     MD 20899-8562}
\author{Y. S. Lee}
\affiliation{Department of Physics and Center for Material
     Science and Engineering, Massachusetts Institute of Technology,
     Cambridge, MA 02139}
\author{P. Smeibidl}
\author{P. Vorderwisch}
\affiliation{4 BENSC, Hahn-Meitner Institute, D-14109 Berlin,
     Germany}
\author{S. Wakimoto}
\affiliation{Department of Physics, University of Toronto,
     Toronto, Ontario M5S 1A7, Canada}

\date{\today}

\begin{abstract}
We have measured the enhancement of the static incommensurate
spin-density wave (SDW) order by an applied magnetic field in
stage-4 and stage-6 samples of superconducting La$_2$CuO$_{4+y}$.
We show that the stage-6 La$_2$CuO$_{4+y}$ ($T_c$=32 K) forms
static long-range SDW order with the same wave-vector as that in
the previously studied stage-4 material. We have measured the
field dependence of the SDW magnetic Bragg peaks in both stage-4
and stage-6 materials at fields up to 14.5 T. A recent model of
competing SDW order and superconductivity describes these data
well.
\end{abstract}

\pacs{74.72.Dn, 75.25.+z, 75.30.Fv, 75.50.Ee}

\maketitle

High-transition-temperature superconductors made by doping
La$_2$CuO$_{4}$ have a very rich magnetic phase diagram as a
function of dopant concentration and the nature of the dopants. In
particular, neutron scattering experiments first demonstrated
incommensurate dynamic spin correlations in
La$_{2-x}$Sr$_x$CuO$_4$ for a wide range of
x.\cite{Dynamic_stripes,Yamada_plot} Later, static incommensurate
spin-density wave (SDW) order was discovered in superconducting
(La$_{2-y-x}$Nd$_y$)Sr$_x$CuO$_4$, La$_{2-x}$Sr$_x$CuO$_4$ at
$x\approx 1/8$, and La$_2$CuO$_{4+y}$,\cite{Tranquada,Kimura,YL}
as well as in insulating La$_{2-x}$Sr$_x$CuO$_4$, $0.02 \leq x
\leq 0.05$. \cite{Waki_0.05} These observations stimulated many
experimental and theoretical studies of the nature of the magnetic
correlations and the static SDW order and their interaction with
superconductivity (SC) in doped La$_2$CuO$_4$ and other cuprate
superconductors.

In our recent neutron scattering experiments on
excess-oxygen-doped La$_2$CuO$_{4+y}$ we have shown that an
applied magnetic field serves as a weak perturbation helping to
probe the nature of this interaction.\cite{Khaykovich02} The
results of Ref. \onlinecite{Khaykovich02}, combined with similar
results on La$_{2-x}$Sr$_x$CuO$_4$ for $x=0.12$,\cite{Katano}
$0.16$\cite{Lake} and $0.10$,\cite{Lake2} together with recent
theoretical studies,\cite{Demler,Zhang,KivelsonLee02} have led to
a consistent picture of microscopic coexistence and competition
between SC and SDW orders. Experimentally, the SDW signal
increases when the field is applied below the superconducting
$T_c$. This occurs because the SC order parameter is suppressed in
the presence of a magnetic field, and as a result, the competing
SDW order is enhanced.\cite{Zaanen,Demler} Additional evidence
that it is the microscopic interaction between SC and SDW order
that drives the enhancement of the SDW peak intensity comes from
recent experiments on lightly-doped non-superconducting
La$_{2-x}$Sr$_x$CuO$_4$. There, an applied field actually {\em
suppresses} the static incommensurate SDW order, in contrast to
the behavior observed in the SC samples.\cite{Matsuda02}

Although there is good qualitative agreement between our previous
experimental results\cite{Khaykovich02} and theory,\cite{Demler} a
quantitative comparison has been difficult to make. This is
primarily because the previous measurements have been limited to
fields below 9 T and by the availability of samples of only one
doping.

Therefore, we have expanded our magnetic field studies of
superconducting La$_2$CuO$_{4+y}$ in two directions. First, we
have applied higher magnetic fields in order to test the predicted
field dependence of the SDW peak intensity. Second, we have
measured the effect of an applied field on two samples of
different doping, a predominantly stage-6 sample (Sample 1) with
$T_c$ = 32 K and a stage-4 sample (Sample 2) with $T_c$ = 42 K.
Magnetic properties of this crystal have been reported in Ref.
\onlinecite{Khaykovich02}, where it is also called Sample 2. We
find that the stage-6 sample exhibits static SDW order similar to
that of the stage-4 sample. We also find that the field dependence
of the SDW Bragg peak intensity, predicted by Demler {\it et
al.},\cite{Demler} describes our results for both samples quite
well. This is especially clear for the stage-4 sample.

There is a significant difference, however, in the relative
increase of the magnetic signal with field for the two samples. We
interpret the difference as arising from different volume
fractions that are ordered magnetically. Based on $\mu$SR
measurements,\cite{Savici,Uemura} it has been argued that the SDW
and SC order parameters are not microscopically homogeneous
throughout the sample, but that small regions of fully developed
static magnetic moment percolate, leading to long-range magnetic
order. It is thus hypothesized that the enhancement of the static
magnetism in an applied field comes primarily from regions where
the SDW order is weak. Therefore, any change in the SDW signal
must depend of the volume fraction of the magnetically ordered
phase at zero field.\cite{Savici,Uemura,Khaykovich02}

\begin{figure}
\centering
\includegraphics[width=3.5in]{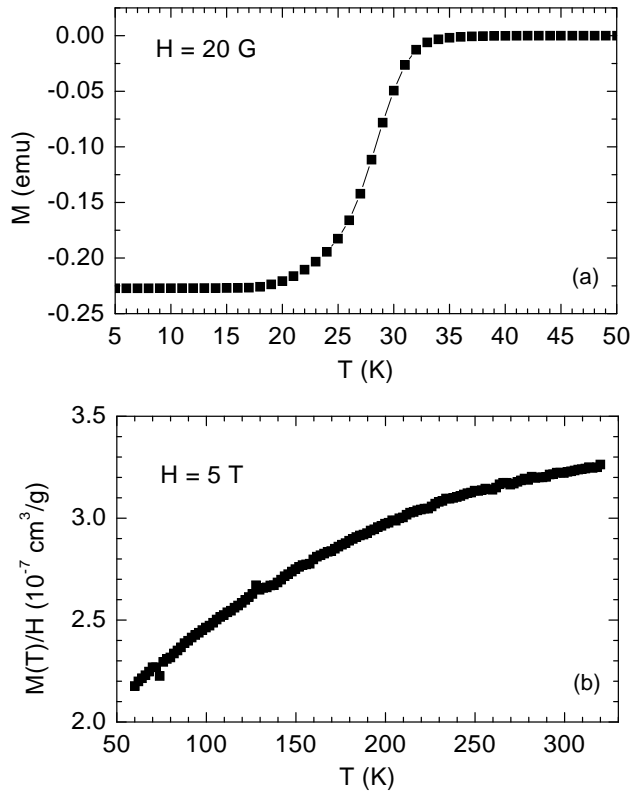}
\vspace{0in}%
\caption{\label{Magnetization}Temperature dependence of the
    magnetization of a small piece of Sample 1. (a) Magnetization showing
    the onset of superconductivity, H = 20 G. (b) Susceptibility showing
    the absence of the weak ferromagnetism, H = 5 T. Similar measurements
    for Sample 2 may be found in Ref. \onlinecite{Khaykovich02}}
\end{figure}

Single crystals of La$_2$CuO$_{4}$ have been grown by the
travelling solvent floating zone technique and subsequently
oxidized in an electrochemical cell, as described
previously.\cite{YL} Sample 2 is prepared by oxidation at room
temperature, while Sample 1 is oxidized at 50 $^\circ$C. SQUID
magnetization measurements have been made on a small piece of each
of the two samples used for the neutron measurements. For Sample 2
(stage-4) the magnetization studies evince a sharp single
transition to the SC state at $T_c =42$ K (onset). For Sample 1
(stage-6), a slightly rounded transition at 32 K (decrease by
10\%) is found. The magnetization of Sample 1 is shown in the Fig.
\ref{Magnetization}a. Magnetization studies also reveal the
absence of weak ferromagnetism, indicating little if any remnant
undoped La${_2}$CuO$_4$. This is shown in Fig.
\ref{Magnetization}b for Sample 1; similar data for Sample 2 have
been published earlier.\cite{Khaykovich02}

Neutron scattering measurements were made on the two crystals,
Sample 1 of 2.9 grams and Sample 2 of 2.6 grams in mass. Our
previously reported measurements of staging behavior, SDW
order,~\cite{YL} and the effect of magnetic
field\cite{Khaykovich02} were made using Sample 2 in magnetic
fields up to 7 T. Elastic neutron scattering studies were
performed at the NIST Center for Neutron Research in Gaithersburg,
MD and at the Hahn-Meitner Institute (HMI), Berlin, Germany. We
used the BT7 thermal triple-axis spectrometer at NIST, with an
incident neutron energy of 13.4 meV, to measure the stage-order of
interstitial oxygen in the Sample 1. High magnetic field
measurements were done on the FLEX cold-neutron triple-axis
spectrometer at HMI with an incident neutron energy of 5 meV. A
pyrolytic graphite (PG) monochromator and a PG analyzer were used,
as well as a PG (for 13.4 meV neutrons) or cold Be (for 5 meV
neutrons) filter to remove higher energy contamination from the
incident neutron beam. The magnetic field was applied using a 14.5
T split-coil superconducting magnet.

The crystal structure of La$_2$CuO$_{4}$ is orthorhombic, space
group {\em Bmab}. The orthorhombicity results from a slight tilt
of the CuO$_6$ octahedra. We therefore use the orthorhombic unit
cell notation with the {\em a-} and {\em b-}axes along the
diagonals of the CuO$_2$ squares and the {\em c}-axis
perpendicular to the layers. In reciprocal space the {\em H, K}
and {\em L} axes are parallel to the {\em a,b}, and {\em c}-axes,
respectively. Excess oxygen in La$_2$CuO$_{4+y}$ is intercalated
between the CuO$_2$ planes, resulting in a mass-density-wave
modulation of the intercalated oxygen density along the {\em
c}-axis. The tilt angle of the CuO$_6$ octahedra changes sign
across the planes of maximum oxygen density, so the tilt reversal
occurs every {\em n}th CuO$_2$ layer. The sample with tilt
reversal every {\em n}th layer is called stage-{\em
n}.~\cite{YL,Wells} The tilt reversals of the CuO$_6$ octahedra
result in structural Bragg peaks displaced along the {\em L} axis
in reciprocal space by $1/n$ from the superlattice Bragg peak.
Because of the staging order, the oxygen doping can not be changed
continuously, and there are miscibility gaps between the lightly
doped antiferromagnet, stage-6, and stage-4 phases.~\cite{Wells}
Samples often consist of a mixture of two or more staged phases.

\begin{figure}
\centering
\includegraphics[width=3.5in]{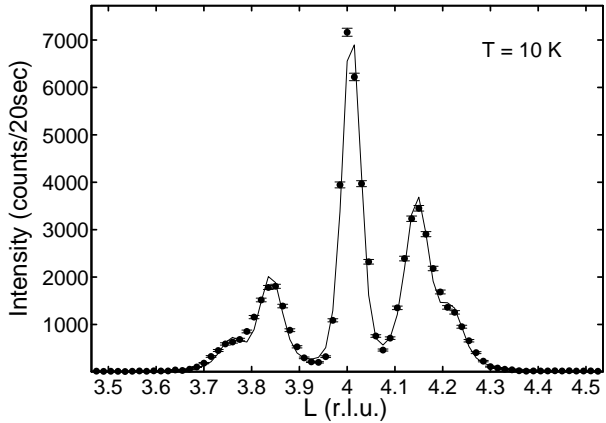}
\vspace{0in}%
\caption{\label{Stage6}Elastic scan across the staging
     superlattice position (0, 1, 4) at low temperatures
     for the stage-6 Sample 1. T = 10 K, $E_i=13.41$ meV, collimation
     30'-30'-S-30'-open, 3-axis mode (S denotes the sample). The fit is
     the result of the convolution of a one-dimensional Lorentzian with
     the instrumental resolution. The two pairs of peaks displaced
     symmetrically around the (014) position result from the staging
     order. The peak at (014) is the {\em Bmab} peak from undoped
     La$_2$CuO$_4$ inclusions.}
\end{figure}

Fig. \ref{Stage6} shows elastic neutron scattering data from the
Sample 1, scanned around the (0,1,4) position in reciprocal space.
The pair of stage-6 peaks (0,1,$4\pm 1/6$) is the strongest, but
the stage-4 peaks at (0,1,$4\pm 1/4$) are also present and
correspond to 20\% of the total intensity of the staging peaks.
The (0,1,4) central peak arises from lightly doped La$_2$CuO$_{4}$
inclusions. In lightly doped samples, this structural Bragg peak
arises from three-dimensional ordering of the CuO$_6$ octahedra
tilts. The intensity of this peak is several orders of magnitude
smaller than that for an undoped La$_2$CuO$_{4}$ sample of similar
size, implying that the undoped fraction is very small. The peaks
are fitted by one-dimensional Lorenzians convoluted with the
instrumental resolution. The staging peaks are almost
resolution-limited, corresponding to a correlation length
exceeding 800 $\AA$. We therefore conclude that this sample
consists of a mixture of stage-4 and stage-6 phases, with the
stage-6 phase dominant. This is in agreement with the results of
magnetic susceptibility measurements, which indicate a lower $T_c$
= 32 K, than a pure stage-4 sample, as noted above. Sample 2,
which has been used in our previous studies, consists
predominately of the stage-4 phase.~\cite{YL,Khaykovich02}

\begin{figure}
\centering
\includegraphics[width=3.5in]{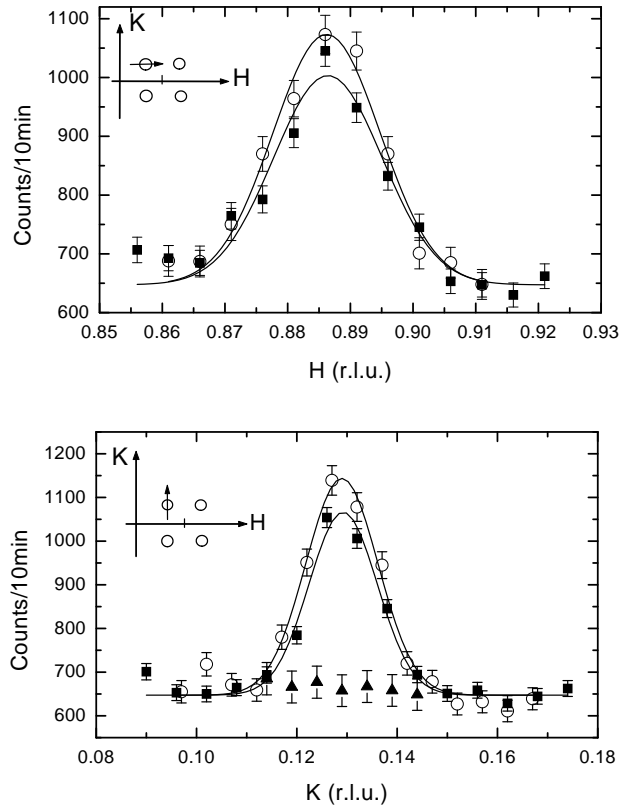}
\vspace{0in} %
\caption{\label{Bragg}Elastic scans through the incommensurate SDW
    Bragg peak at {\bf Q} = (0.886, 0.129, 0) at zero applied field
    (squares) and H = 14.5 T (circles) for the stage-6 Sample 1.
    The sample is oriented such that the neutron wave vector transfer Q
    is parallel to the CuO$_2$ planes ({\it (H, K, 0)} in reciprocal space).
    The applied field is perpendicular to the CuO$_2$ planes. Measurements
    are made at $T = 3$ K and the background (triangles) is measured at
    $T = 45$ K. The collimation is
    60'-S-60'-open, $E_i=5$ meV. The inset shows schematically the positions of 4
    incommensurate SDW peaks in reciprocal space together with the scan
    directions. The line-profile is Gaussian; the HWHM is
    resolution-limited.}
\end{figure}

Static long-range incommensurate SDW order has been found in the
stage-4 sample below $T_m=42$ K by neutron scattering.\cite{YL} In
previous work we have shown that a magnetic field, applied
perpendicular to the CuO$_2$ layers, increases the intensity of
the incommensurate magnetic Bragg peaks approximately linearly
with field for fields up to at least 7 T. Here we find that
similar behavior occurs for the stage-6 sample. Fig. \ref{Bragg}
shows the magnetic Bragg peak resulting from the incommensurate
SDW order at $H=0$ and $14.5$ T, in Sample 1. We have checked that
the peak positions at ($1\pm 0.114,0\pm 0.128,0$) are identical
for both Sample 1 and Sample 2 ((1, 0, 0) is the position of the
antiferromagnetic zone center for undoped La$_2$CuO$_4$). Fig.
\ref{Bragg} clearly shows that only the intensity of the SDW peak
increases with field, whereas the width and the position of the
peak remain the same. The peak widths are limited by our
instrument resolution, which indicates long-range magnetic order
with the correlation length in excess of 400 $\AA$.

\begin{figure}
\centering
\includegraphics[width=3.5in]{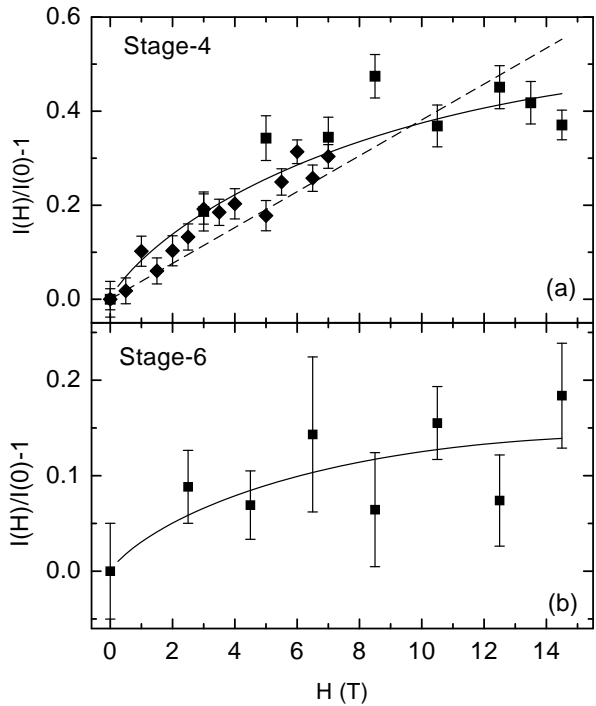}
\vspace{0in} %
\caption{\label{FieldDep}Field dependence of the SDW Bragg peak
    intensity I(H), normalized to the zero-field intensity, I(0).
    (a) Sample 2, stage-4. The diamonds correspond to data taken
    at NIST and reported previously.\cite{Khaykovich02} The
    squares represent the data taken at HMI. (b) Sample 1,
    stage-6, measured at HMI. The solid lines correspond to the
    function $(H/H_{c2})\ln(\theta H_{c2}/H)$ and the dashed line
    corresponds to a linear function $H/H_{c2}$, as discussed in the
    text. The larger error bars and smaller number of points for the
    HMI experiment result from limited access to the instruments.
    }
\end{figure}

Fig. \ref{FieldDep} shows the field dependence of the magnetic
Bragg peak intensity for Sample 2 (a) and Sample 1 (b). The
intensity is normalized to the zero-field value. Fig.
\ref{FieldDep}(a) shows data taken both at NIST (below 7 T) and at
HMI (0 T to 14.5 T). The points at 0, 3, 5 and 7 T have been
measured on both spectrometers. The intensity I(H) is obtained by
averaging counts from both longitudinal and transverse scans close
to the peak position. The field-independent background has been
measured and subtracted.

One of the most intriguing properties of stage-4 oxygen-doped
La$_2$CuO$_{4+y}$, reported in Ref. \onlinecite{YL}, is that the
onset of the SDW Bragg peak coincides with the superconducting
$T_c = 42$ K. Therefore, we have measured the temperature
dependence of the SDW Bragg peak for Sample 1, as shown in Fig. 5.
Clearly, the SDW signal appears first at $T_m = (40\pm2)$ K, well
above $T_c = 32$ K (see Fig. \ref{Magnetization}a). However, since
Sample 2 contains about 20\% of the stage-4 phase, this onset of
the magnetic order at $40\pm2$ K may arise from the ordering of
the stage-4 part alone. A closer look at the magnetization curve
in Fig. \ref{Magnetization}a reveals a small drop of the
magnetization at 42 K. We ascribe this to the superconducting
transition of the stage-4 phase. To test whether $T_c$ and $T_m$
are the same for stage-6 alone, one would need a homogeneous
stage-6 sample.

\begin{figure}
\centering
\includegraphics[width=4in]{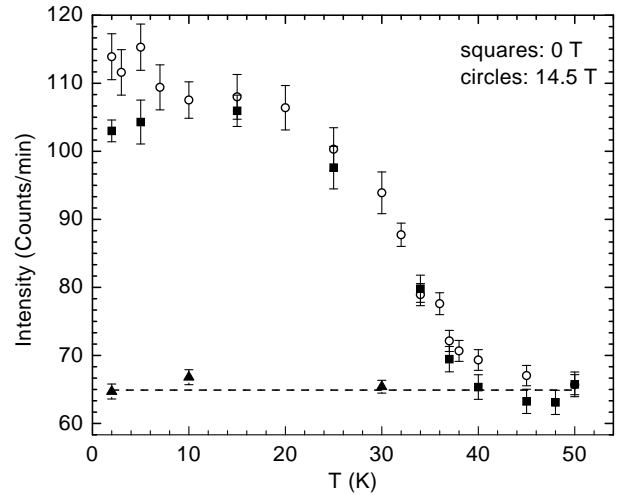}
\vspace{0in} %
\caption{\label{TempDep}Temperature dependence of the SDW Bragg
    peak intensity, Sample 1, stage-6. $H$ = 0 T (squares) and 14.5
    T (circles). The triangles correspond to the background
    determined by fitting SDW data, like these in Fig. 3, to a
    Gaussian peak plus a $\bf{Q}$-independent background.
    At $T = 2$ K, the background comes from a scan at $H = 0$ T and
    at the higher temperatures from scans at $H = 14.5$ T.
    The dashed line is a linear fit to the background points;
    the background is temperature- and field-independent.
    }
\end{figure}

We now discuss these experimental results. Although Sample 1
consists of a mixture of stage-6 and stage-4 phases, the two
phases must have identical SDW ordering at low temperatures for
the following reasons: Samples 1 and 2 are of similar size, and
their incommensurate SDW peaks have intensities of ($40\pm2$) and
($30\pm2$) counts/min, respectively, at zero field. However, only
20\% of Sample 1 is in the stage-4 phase. Were only the stage-4
phase magnetically ordered, the SDW peak intensity from Sample 1
would be smaller than that from Sample 2, even if 100\% of the
stage-4 phase were magnetically ordered. It is also important to
note that the incommensurability must be identical for stage-4 and
stage-6. This is evident from the sharp, resolution-limited SDW
peaks. Were the peaks for stage-4 and stage-6 shifted relative to
each other, a wide SDW peak would result. Finally, a pure stage-6
sample has been measured recently,\cite{YLunpublished} and this
sample shows sharp SDW peaks at the same positions. Unfortunately,
this pure stage-6 crystal is too small for quantitative
field-dependence measurements. The single value of the
incommensurability $\delta$ observed in the stage-4 and stage-6
samples may result from the saturation of $\delta$ with hole
concentration. For the inelastic scattering in Sr-doped
La$_{2-x}$Sr$_x$CuO$_4$, $\delta$ is equal to $x$ for small $x$,
but saturates above $x\simeq0.12$.\cite{Yamada_plot} The similar
saturation of $\delta$ is seen for the SDW in Nd codoped
samples.\cite{Tranquada} As shown in Ref.
\onlinecite{Khaykovich02}, the doping of Sample 2 corresponds to a
hole concentration of 0.14 per Cu, above the concentration
necessary for saturation. Stage-6 samples probably also have hole
concentration high enough for saturation. The incommensurability
is close to the saturation value in the other La$_2$CuO$_4$-based
materials. Although the saturation can thus explain quite similar
values of $\delta$ in stage-4 and stage-6 samples, it is
nonetheless surprising that they are identical within the
experimental error of 0.002 (r.l.u.).

The model of competing order parameters\cite{Demler,KivelsonLee02}
predicts the existence of three different phases: a pure SC phase,
a SDW+SC phase, in which SDW order coexists with and couples to
the SC order, and a pure SDW phase. The transitions between these
phases occur by changing a control parameter, that is related to
doping, or applied magnetic field. At zero magnetic field, stage-4
La$_2$CuO$_{4+y}$ apparently resides in the SDW+SC phase. By
increasing the doping, La$_2$CuO$_{4+y}$ is expected to move into
the pure SC phase. Unfortunately, it does not appear to be
possible to increase the doping beyond the value that gives
stage-4 by additional oxidation of La$_2$CuO$_{4+y}$. (This can,
of course, be accomplished by changing $x$ in
La$_{2-x}$Sr$_x$CuO$_{4}$\cite{Katano,Lake,Lake2}). When the
doping decreases, La$_2$CuO$_{4+y}$ is expected to remain in the
SDW+SC phase, and this is confirmed by our observations.

The enhancement of the SDW order parameter in an applied field
originates from the suppression of superconductivity near
vortices. The superconducting order parameter $|\psi|^2$ is most
strongly suppressed inside the vortex core, of radius $\xi\simeq
30$ $\AA$. But $|\psi|^2$ is also suppressed below its zero-field
value even far from the cores, recovering as ($1-2\xi^2/r^2$).
Here $\xi$ is the superconducting correlation length and $r$ is
the distance from the core. Since the magnetic correlation length
exceeds 400 $\AA$, which is much larger than $\xi$, averaging
$|\psi(r)|^2$ over the regions far away from the vortex is
required to account for the resulting change of the SDW order
parameter. The SDW peak intensity is thus predicted to increase
with field as $\Delta I \sim (H/H_{c2})\ln(\theta H_{c2}/H)$; the
logarithmic correction originates from the $1/r$ term in the
suppression of $|\psi|^2$. Here $H_{c2}$ is the upper critical
field of the superconductor, and $\theta$ is a constant of order
unity, independent of doping.\cite{Demler}

We have fitted the data in Fig. \ref{FieldDep} using this
prediction. The best fit gives $H_{c2}=49$ T for the stage-6
Sample 1 and 67 T for the stage-4 Sample 2. In both cases $\theta$
is set to 1 for the best fit. We also show the result of a fit to
a simple linear dependence, $H/H_{c2}$, for Sample 2 (dashed line
in Fig. \ref{FieldDep}a), which describes our previous low field
results adequately.\cite{Khaykovich02} Clearly, the logarithmic
correction to the linear dependence improves the agreement between
the theoretical prediction and the data. The high field data
points above 7 T are especially important in order to distinguish
between the two possibilities. The ``goodness-of-fit'', $\chi^2$,
analysis confirms this conclusion: $\chi^2 \simeq 0.002$ is
significantly smaller for the $H\ln H$ fit, than $\chi^2 \simeq
0.006$ for the linear fit.

Numerical calculations have resulted in $\theta \approx 3$ for a
triangular vortex lattice.\cite{Demler} Our experimental results
suggest that $\theta$ is smaller, probably because the vortex
lattice in La$_2$CuO$_{4+y}$ is highly disordered. Further, by
analogy with recent results on La$_{1.83}$Sr$_{0.17}$CuO$_4$, the
vortex lattice is, in all likelihood, square rather than
triangular.\cite{vlattice}

The upper critical fields obtained from fitting our data are in
reasonable agreement with the commonly accepted values for doped
La$_2$CuO$_4$-based superconductors. The applied field of 14.5 T
is, therefore, much smaller than the upper critical field at T = 2
K, the temperature of the samples during the neutron scattering
experiments. Therefore, the applied magnetic field acts only as a
weak perturbation of the superconducting state. We have previously
reported that the resistive superconducting transition of the
stage-4 sample is shifted significantly to lower temperature by
applied fields up to 5 T.\cite{Khaykovich02} Similar measurements
have not been done for the stage-6 sample, because putting
electrical contacts on these heavily oxidized samples is very
difficult. We point out, however, that resistance does not provide
a thermodynamic measurement of the superconducting phase
transition, as does, for example, the specific heat.

Fig. \ref{FieldDep}b shows the enhancement of the SDW peak
intensity with field for Sample 1. The effect of applied field is
much smaller for this sample, compared with the stage-4 one. The
smaller change of the peak intensity may be explained by a model
in which the magnetic volume fraction is less than unity, as
discussed in Ref.~\onlinecite{Khaykovich02}. $\mu$SR measurements
on a piece of Sample 2 indicate that it consists of small magnetic
clusters, in which the incommensurate order has its full local
moment, and that the rest of the material is
non-magnetic.\cite{Savici,Uemura} Furthermore, $\mu$SR
measurements indicate that the fraction that is magnetically
ordered is $F\simeq 0.4$. According to the model, the
superconducting order parameter is suppressed in the
magnetically-ordered regions even at zero applied field. Then the
enhancement of the magnetic order in the field originates from
predominately non-magnetic regions, where the superconductivity is
strong at zero field. Therefore, the increase in the SDW peak
intensity $\Delta I \sim (1-F)/F$. Comparing the relative increase
of intensity with field for Samples 1 and 2, we conclude that
about 60\% of Sample 1 is magnetically ordered at zero field.
$\mu$SR measurements on the stage-6 sample would serve to test
this conclusion.

In conclusion, we have studied the effects of an applied magnetic
field on the static incommensurate SDW order in excess-oxygen
doped La$_2$CuO$_{4+y}$. We have shown that the SDW order exists
in both stage-4 and stage-6 materials. The magnetic ordering
wave-vector is exactly the same for both dopings. By applying a
high magnetic field of up to 14.5 T, we measure the field-induced
change in the SDW Bragg peak intensity. The field-induced
enhancement of the SDW signal follows the prediction of the
recently proposed model of competing superconducting and magnetic
orders in doped La$_2$CuO$_{4}$. The $H\ln H$ field dependence of
the SDW signal demonstrates the quantitative agreement between
experiment and theory.

\begin{acknowledgments}
We have benefited from useful discussions with E. Demler, S.
Sachdev, P. A. Lee and Y. J. Uemura. This work has been supported
at MIT by the MRSEC Program of the National Science foundation
under Award No. DMR 02-13282, by NSF under Awards No. DMR 0071256
and DMR 99-71264. Work at the University of Toronto is part of the
Canadian Institute for Advanced Research and is supported by the
Natural Science and Engineering Research Council of Canada. We
acknowledge the support of the National Institute of Standards and
Technology, U.S. Department of Commerce, and the Berlin Neutron
Scattering Center, Hahn-Meitner Institute, Berlin, in providing
the neutron facilities used in this work.
\end{acknowledgments}


\end{document}